\documentclass[pdftex,twocolumn,epjc3]{svjour3}
\journalname{Eur. Phys. J. C}

\usepackage[linktocpage,unicode]{hyperref}
\usepackage{csquotes,graphicx,fancyhdr}
\usepackage{setspace,longtable,graphicx,hyphenat,hyperref,fancyhdr,ifthen,enumitem}
\usepackage[dvipsnames]{xcolor}
\usepackage[export]{adjustbox}
\hypersetup{unicode=true,colorlinks=true,linkcolor=black,citecolor=blue,urlcolor=blue}
\usepackage{url}
\usepackage{indentfirst}
\usepackage{amssymb,amsmath}
\usepackage{geometry}
\usepackage[style=gost-numeric,bibencoding=auto,backend=biber,sorting=none,language=auto,autolang=other,doi=false,url=false]{biblatex}
\AtEveryBibitem{\clearfield{day}
	\clearfield{month}
	\clearfield{endday}
	\clearfield{endmonth}}
\bibliography{Ru-Article_6}
\usepackage{multido}
\newcommand{\e}{\text{e}}
\newcommand{\n}{\text{n}}
\newcommand{\D}{\text{D}}
\newcommand{\M}{\text{M}}
\newcommand{\N}{\text{N}}
\newcommand{\K}{\text{K}}

\newcommand{\CC}{\text{C}}
\newcommand{\LL}{\text{L}}
\newcommand{\HH}{\text{H}}
\newcommand{\Pl}{\text{Pl}}
\newcommand{\cc}{\text{c}}
\makeatletter
\ams@newcommand{\vardot}[2]{%
  {\mathop{#2\kern0pt}\limits^{\vbox to-1.4\ex@{\kern-\tw@\ex@
   \hbox{\normalfont\multido{}{#1}{.}}\vss}}}}
 \makeatletter
\def\@fnsymbol#1{\ensuremath{\ifcase#1\or \dagger\or  *\or \dagger\dagger
   \or \ddagger\ddagger \else\@ctrerr\fi}}
    \makeatother
\makeatother

\begin{document}

\title{Flexible extra dimensions}

\author{{Polina Petriakova}
\thanksref{e1,addr1}
\and {Arkady~A.~Popov}
\thanksref{e2,addr2}
\and
{Sergey~G.~Rubin}
\thanksref{e3,addr1,addr2}
}

\thankstext{e1}{e-mail: plinapolina@yandex.ru}
\thankstext{e2}{arkady\_popov@mail.ru}
\thankstext{e3}{sergeirubin@list.ru}

\institute{National Research Nuclear University MEPhI (Moscow Engineering Physics Institute), 115409, Kashirskoe shosse 31, Moscow, Russia \label{addr1}
          \and
          N.~I.~Lobachevsky Institute of Mathematics and Mechanics, Kazan  Federal  University, 420008, Kremlevskaya  street  18,  Kazan,  Russia  \label{addr2}
}

\date{Received: date / Accepted: date}

\maketitle

\begin{abstract}
This paper discusses the origin of the small parameters with the aim of explaining the Hierarchy problem. The flexible extra dimensions are an essential tool in the process by which physical parameters are formed. The evolution of a multidimensional metric starts at the Planck scale and is completed with the static extra-dimensional metric and the 4-dim de Sitter space at high energies, where the exponential production of causally disconnected universes begins. Quantum fluctuations independently distort the metric within these universes, causing inflationary processes within them. Some of these universes tend asymptotically towards states characterised by small Hubble parameters.  The effective parameter reduction applied to the Higgs sector of the Standard Model is explained by the presence of small-amplitude distributions of a scalar field in a fraction of these universes. 
\end{abstract}

\section{Introduction}

The success of the inflationary paradigm in describing the early universe clearly suggests that the physical laws are formed at high energies, where we can only guess at the Lagrangian structure \cite{Brandenberger:2006vv,Tegmark:2005dy}. Evidently, the physics is already formed at least on the inflationary scale $E_I \sim 10^{13}$ GeV and the observable low energy physics depends on parameters and initial conditions formed at high energies  \cite{Loeb:2006en,Ashoorioon:2013eia}.
The extra dimensions have now become a widely used tool to obtain new theoretical results \cite{Abbott:1984ba,Chaichian:2000az,Randall:1999vf,Brown:2013fba,Bronnikov:2009zza}. It is reasonable to suppose that their metric $g_\n$ is also formed at the high energy scale. Here we make use the idea of flexible (inhomogeneous) extra space that has been developed in \cite{Gani:2014lka,Rubin:2015pqa,Rubin:2014ffa}.

Reduction of a multidimensional action to an observed 4-dimensional one
\begin{equation}\label{SDto4}
   S_{4+\n}(\lambda_{4+\n},g_\n)\to S_4(\lambda_4[g_\n]) 
\end{equation}
is necessary element in all models with $\n$ extra dimensions. Therefore, 4-dimensional parameters $\lambda_4$ inevitably depend on the extra metric,
\begin{equation}\label{l4}
    \lambda_{4}=\mathcal{Z} (g_{\n},\lambda_{4+\n}),
\end{equation}
and a future theory should take this fact into account. 
A functional $\mathcal{Z}$ is specific to any extra dimensional model with initial parameters $\lambda_{4+\n}$. 
By solving the system of equations \eqref{l4} with appropriate precision, one could determine the primary parameters $\lambda_{4+\n}(M)$ at a chosen scale $M$.

The situation is much more complex when one considers the smallness of the observable parameters below the electroweak scale compared to the parameter values at the Planck scale, $\lambda_4 \lll \lambda_{4+\n}$. Nevertheless, activity in this direction has been observed. The paper \cite{Krause:2000uj} uses warped geometry to solve the small cosmological constant problem. The hybrid inflation \cite{2002PhRvD..65j5022G} was developed to avoid the smallness of the inflaton mass. The electron to proton mass ratio is discussed in \cite{Trinhammer:2013mxa}. The seesaw mechanism is usually used to explain the smallness of the neutrino to electron mass ratio \cite{Fujikawa:2016ocu}. The implementation of this activity in its entirety is a matter for the future. 
	
The aim of this paper is to discuss an application of flexible extra metrics - a set of metrics of the cardinality of the continuum -  to the Hierarchy problem. The essence of the latter is expressed in the question: Why are the observable low energy physical parameters so small as compared to those at the Planck scale. It is implied that the parameters formed at the Planck scale are of the order of this scale which seems natural. It has been shown earlier \cite{Rubin:2015pqa, 2017JCAP...10..001B} that this approach gives encouraging results for explaining of the Cosmological Constant smallness. The effect of the quantum corrections in this content was discussed in \cite{Rubin:2020pqu}. Here we continue this research by including the Higgs sector of the Standard Model. Special attention is paid to the mechanism of the appearance of small parameters and the role of the background fields. 

The idea that a Lagrangian parameters can be considered as some functions of a field has been widely used since Schwinger's paper  \cite{Schwinger:1951nm,}. Such fields can be involved in the classical equations of motion together with the ''main'' fields or treated as background fields. The latter were applied for the fermion localization on branes \cite{Sorkhi:2018nln,Sui:2017gyi,Arai:2018hao}, gauge fields localization \cite{Chumbes:2011zt}, extensions of the gravity in the form $f(\phi)R$ \cite{Bronnikov:2003rf} and so on.

As a mathematical tool, we  use the effective field theory technique - well-known method for theoretical investigation of the energy dependence of physical parameters \cite{Peskin:1995ev}. In this approach, the parameters $\lambda_i (M)$  of the Wilson action are fixed at a high energy scale $M$. The renormalization flow used to descend to low energies (the top-down approach) is discussed in \cite{Burgess:2013ara,Hertzberg:2015bta,Babic:2001vv,Dudas:2005gi}. As is usually stated, the parameters $\lambda_{4+\n} (M)$ of the Wilson action already contain quantum corrections caused by field fluctuations with energies between the chosen scale $M$ and a maximal energy scale. Therefore the natural values of these parameters are typically many orders of magnitude larger than the electroweak scale $v\simeq 100$ GeV. 
	
The research is also based on the multidimensional $f(R)$ gravity. The interest in $f(R)$ theories is motivated by inflationary scenarios starting with the work of Starobinsky \cite{Starobinsky:1980te}. The simplest extension of the gravitational theory is that which includes a function of the Ricci scalar $f(R)$. In the framework of such an extension, many interesting results have been obtained. Some viable $f(R)$ models in 4-dimensional space that satisfy the observational constraints are proposed in Refs. \cite{DeFelice:2010aj,2014JCAP...01..008B,Sokolowski:2007rd,2007PhLB..651..224N,Nojiri_2017,}. 
The stabilization of extra space as a pure gravitational effect has been studied in \cite{2003PhRvD..68d4010G,2002PhRvD..66d4014G}. 
 
In the framework of the scalar-tensor theory, Weinberg \cite{1984NuPhB.237..397C}  has proved that the strong fine-tuning of initial parameters of a Lagrangian is unavoidable if the metric and scalar fields are constant in space-time. The latter implies that the solution of the problem should be sought in the class of non-uniform configurations of metrics and fields. Flexible metrics discussed in this paper belong to this class. 

Our preliminary study of the inhomogeneous extra metrics concerns parameters such as the cosmological constant \cite{Rubin:2015pqa}, parameters of the Starobinsky inflationary model and the baryon asymmetry of the Universe. 
It has been shown there that the inhomogeneous metrics can be tuned to explain the smallness of these effective parameters.
Below we develop of our approach to the Hierarchy problem and involve the Higgs Lagrangian in our consideration.

The metric ansatz used in this paper has been widely studied in the linear gravity \cite{2000PhRvD..62d4014O,2003PhRvD..68b5013C,2005PhRvD..71h4002S,2005PhRvD..71j4018R}, applying in particular to the solution of the Hierarchy problem \cite{2000PhRvL..84.2564G,2000PhRvL..85..240G}.

\section{General relationships}\label{Outlook}

Consider the $f(R)$ theory of gravity with a minimally coupled scalar field $\zeta$ in $\D = 4 + \n$ dimensions 
\begin{eqnarray}\label{S0}
S = \frac{m_{\D}^{\D-2}}{2}  \int_{M_\D} && d^{\D} X \sqrt{|g_{\D}|} \,  \Bigl( f(R) 
\nonumber \\ &&
+ \partial^{\M}\zeta \, \partial_{\M}\zeta -2 V\bigl(\zeta \bigr) \Bigr)\, ,
\end{eqnarray}
where $M_\D$ is $\D$-dimensional manifold, $g_{\D} \equiv \text{det} g_{\M\N}$, $\M,\N =\overline{1,\D}$, 
the $\n$-dimensional  manifold $M_\n$ is assumed to be closed one, $f(R)$ is a function of the D-dimensional  Ricci scalar $R$, and $m_\D$ is the 
$\D$-dimensional Planck mass. In 
the following we set $m_\D=1.$ 

The scalar field $\zeta$ affects the extra space metric through the Einstein equations, but here we are interested in small amplitude solutions of this field, i.e. %
$$\zeta(X)\ll 1.$$
Therefore, its role in the metric formation is negligible and it is considered as a auxiliary or trial field acting in the background metric. This approximation makes the analysis slightly easier, but is not very significant. The importance of this field becomes crucial when we consider its interaction with other fields, such as the proto-Higgs field $H_\text{P}(x)$ in our case. We assume that the auxiliary field $\zeta$ plays the role of the background field introduced by Schwinger (see references in the Introduction). This means that the inequality 
$$H_\text{P}\ll\zeta$$
should be kept in mind.

Variation of action \eqref{S0} with respect to the metric $g^{\M\N}_\D$ and scalar field leads to the known~equations
\begin{align}\label{eqMgravity}
&-\frac{1}{2}{f}(R)\delta_{\N}^{\M} + \Bigl(R_{\N}^{\M} +\nabla^{\M}\nabla_{\N} - \delta_{\N}^{\M} \Box_{\D} \Bigr) {f}_R  \nonumber \\
&= - T_{\N}^{\M}, 
\\ 
\label{eqMscalarfield} &\Box_{\D} \, \zeta + V^{\prime}_{\zeta} =0,  \end{align}
with $f_R = \dfrac{df(R)}{dR}$, $\Box_{\D}= \nabla^{\M} \nabla_{\M}$ and $V^{\prime}_{\zeta} = \dfrac{d V\bigl(\zeta \bigr)}{d\zeta}$. The arbitrary potential satisfies conditions $\left.V \bigl(\zeta \bigr)\right|_{\zeta =0}=0$, $\left. V'_{\zeta} \bigl(\zeta \bigr)\right|_{\zeta =0}=0$. Equation \eqref{eqMscalarfield} is known to be the consequence of equations \eqref{eqMgravity}. The corresponding stress-energy tensor of the scalar field $\zeta$ is
\begin{align}
& T_{\N}^{\M} = \frac{\partial L_{matter}}{\partial\bigl(\partial_{\M} \zeta \bigr)}\partial_{\N} \zeta - \frac{\delta_{\N}^{\M}}{2} L_{matter} \nonumber \\
& = \partial^{\M}\zeta \, \partial_{\N}\zeta - \frac{\delta_{\N}^{\M}}{2} \, \partial^{\K}\zeta \, \partial_{\K}\zeta +  \delta_{\N}^{\M}V\bigl(\zeta \bigr) \, .
\end{align}
We use the following conventions for the curvature tensor $R_{\M\N\K}^\LL=\partial_\K\Gamma_{\M\N}^\LL-\partial_\N \Gamma_{\M\K}^\LL +\Gamma_{\CC\K}^\LL\Gamma_{\N\M}^\CC-\Gamma_{\CC\N}^\LL \Gamma_{\M\K}^\CC$
and the Ricci tensor $R_{\M\N}=R^\K_{\M\K\N}$.

The quantum fluctuations at the de Sitter stage can break the maximally symmetrical extra space metric \cite{2021arXiv210908373R}, which is the reason for an inhomogeneous metric formation \cite{Rubin:2015pqa}. Here we consider an inhomogeneous n-dimensional extra metric
\begin{align}\label{metric_deformed_r}
ds^{2} = &g_{\M\N}X^\M X^\N
=\e^{2\gamma(u)}\Bigl(dt^2 - \e^{2Ht}\delta_{ij}dx^i dx^j\Bigr) 
\nonumber \\ &
- du^2 - r^2(u)\,d\Omega_{\n-1}^2 \, , \quad i,j=\overline{1,3}
\end{align}
with the coordinates $X^4 \equiv u$ and $X^{\mu}\equiv x^{\mu}, \mu=0,1,2,3.$
The Hubble parameter $H$ and the static metric function $r(u)$ are solutions of the system below. The Ricci scalar
\begin{align}\label{Ricci_n_dim_deformed_r}
& R(u)= 12 H^2 \e^{-2 \gamma(u)} -  8 \gamma'' - 20 {\gamma'}^2 - \bigl(\n-1 \bigr) \left( \dfrac{2 r''}{r}\right. \nonumber \\ 
&  \left.+ \dfrac{8 \gamma' r^{\prime}}{r}\,+ \bigl(\n-2 \bigr) \left(\dfrac{r^\prime}{r}\right)^2\,- \dfrac{(\n-2)}{r^2} \right)
\end{align}
does not depend on time. The notation used are ${}' \equiv d / du$ and ${}'' \equiv d^2 / du^2$ respectively.
Then system \eqref{eqMgravity} for $(tt)=...=(x^3 x^3)$, $(uu)$ and $(x^5 x^5)=...=(x^{\D-1} x^{\D-1})${--}components and \eqref{eqMscalarfield} become
\begin{align}\label{eq_tt_deformed_r}
& {R'}^2 f_{RRR} +\left(R'' +3 \gamma' R' + \bigl(\n-1 \bigr) \frac{r'}{r} R' \right)f_{RR} \nonumber \\ 
& - \left( \gamma'' +4{\gamma'}^2 + \bigl(\n-1 \bigr)\frac{\gamma' r'}{r} - 3H^2\e^{-2 \gamma(u)} \right)  f_{R} \nonumber \\ 
&- \dfrac{f(R)}{2} = - \dfrac{{\zeta'}^2}{2}   - V\bigl( \zeta \bigr), \\
\label{eq_thetatheta_deformed_r}
& \left( 4\gamma'R' + \bigl(\n-1 \bigr) \dfrac{r'}{r} R' \right)\,f_{RR} - \left( 4 \gamma'' + 4{\gamma'}^2 
\right. \nonumber \\
& \left.+ \bigl(\n-1 \bigr) \dfrac{r''}{r} \right) \, f_R - \, \dfrac{f(R)}{2} =   \dfrac{{\zeta'}^2}{2}  - V\bigl(\zeta \bigr) \,, \\
\label{eq_phiphi_deformed_r}
& {R'}^2 f_{RRR} \, + \biggr( R'' + 4\gamma' R' + \bigl(\n - 2\bigr) \dfrac{r'}{r}\,R'  \biggl) f_{RR} \nonumber  \\  & \quad  - \biggl(\dfrac{r''}{r} + \frac{4\gamma' r'}{r} + \bigl(\n-2 \bigr)\dfrac{{r'}^2}{r^2} 
 \nonumber \\ &
- \dfrac{(\n-2)}{r^2} \biggr) f_R  - \, \dfrac{f(R)}{2} = -  \dfrac{{\zeta'}^2}{2}  - V\bigl(\zeta \bigr) \, , \\
\label{eq_scalar_deformed_r}
& \zeta'' + \left(4 \gamma' + \bigl(\n-1\bigr)\dfrac{r'}{r}\right) \, \zeta'  - V^{\prime}_{\zeta} =0.
\end{align} 
It can be shown that one of these equations is a consequence of the others. 
Also, we will use the definition of the Ricci scalar
\eqref{Ricci_n_dim_deformed_r}
as the additional equation for $R(u)$ which will be treated as a new unknown function to avoid  3rd and 4th order derivatives in the equations.

The combination $2\cdot$\eqref{eq_thetatheta_deformed_r}$-f_R \cdot$\eqref{Ricci_n_dim_deformed_r} is constraint equation  \begin{align}\label{cons}
&\biggr(8\gamma'R' + 2\bigl(\n-1\bigr)\dfrac{r'}{r}R' \biggl) f_{RR} + \Biggl(12 {\gamma'}^2 
\nonumber \\ &
+\bigl(\n-1\bigr) \biggl(\dfrac{8\gamma' r'}{r} + \bigl(\n-2\bigr)\dfrac{\bigl({r'}^2-1\bigr)}{r^2} \biggr) +R \Biggr)f_R \nonumber \\ 
& - 12H^2\e^{-2 \gamma(u)} f_R\, - f(R) =  {\zeta'}^2 - 2 V\bigl( \zeta \bigr)	
\end{align}
containing only first-order derivatives. It plays the role of a restriction on the solutions of the coupled second-order differential equations.

\section{Static extra dimensional metrics }

\begin{figure}[!th]
\centering
\includegraphics[width=0.4\linewidth]{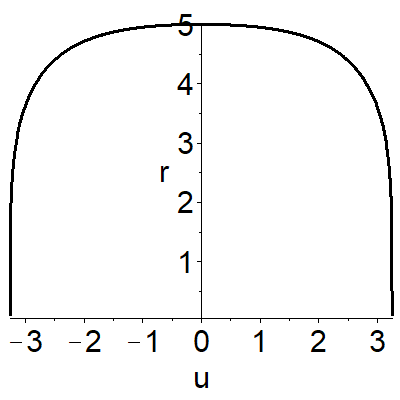} \qquad
\includegraphics[width=0.4\linewidth]{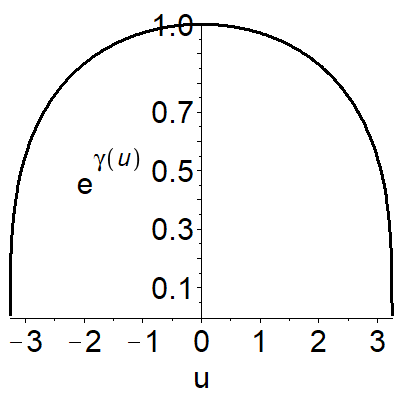} \qquad
\\ \includegraphics[width=0.4\linewidth]{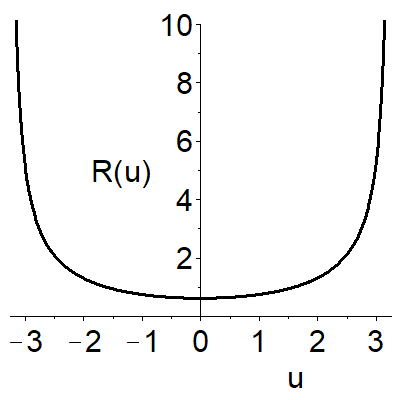}  \qquad
\includegraphics[width=0.4\linewidth]{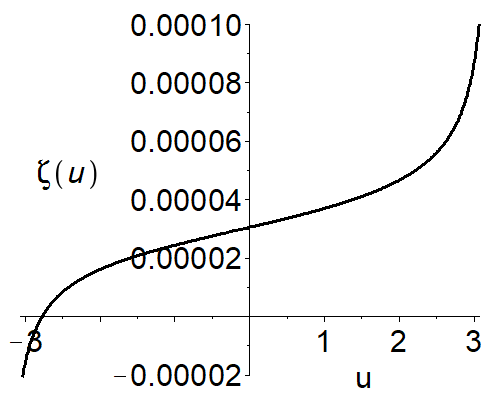} \qquad 
\caption{Solution for $m_\D=1, \n=3, a=1$, $c=0.2$, $H=0$, $V(\zeta) = m_{\zeta}^2 \zeta^2 /2$, $ m_{\zeta}=0.1$, $r(0)=5$, $r'(0)=0$, $\gamma(0)=0$,  $\gamma'(0)=0$, $R(0) \simeq 0.62$ (from equation \eqref{cons}), $R'(0) = 0$, $\zeta(0) \simeq 3.04 \cdot 10^{-5}$, $\zeta'(0) = 6 \cdot 10^{-6}$, $u_{\text{max}}=-u_{\text{min}} \simeq 3.25$, $r(u_{\text{min}})=r(u_{\text{max}})=0 $, $m_{\Pl} \simeq 65.9 m_\D$. 
}
\label{11}
\end{figure}

The form of stationary extra space metric is the result of a metric evolution governed by the classical equations of motion, and hence depends on initial configurations. One can imagine an analogy with the Schwarzschild metric which explicitly depends on an initial matter distribution. The system  \eqref{Ricci_n_dim_deformed_r}-\eqref{eq_scalar_deformed_r} is the set of second order derivative equations and hence must be endowed by additional conditions if one wish to obtain a particular solution. To this end, the metric functions $\gamma(u), r(u), R(u)$ and the matter field $\zeta(u)$ could be fixed at a certain point $u_c$ together with their first derivatives. These conditions are linked by constraint \eqref{cons}. For example, the set
\begin{align}\label{cond_2branes}
&r(u_\cc)=r_0, \   R(u_\cc)=R_0,  \   \gamma(u_\cc)=0, \  \zeta(u_\cc) = \zeta_0, \nonumber \\
& r^\prime(u_\cc)=R^\prime(u_\cc) = \gamma ^\prime(u_\cc) = \zeta^\prime(u_\cc)= 0,
\end{align}
at $u_\cc = 0$ leads to  symmetrical distributions. One of the solution is shown in Figs.\ref{11} where the auxiliary field is extremely small due to the choice $\zeta(u_\cc)\ll 1, \zeta^\prime(u_\cc)\ll 1$. The smallness of the scalar field value (Fig.\ref{11}, right panel) may cause concern. We discuss the stability of such solutions and their possible breaking under the quantum fluctuations in the next subsection.

\begin{figure}[!th]
\centering
\includegraphics[width=0.4\linewidth]{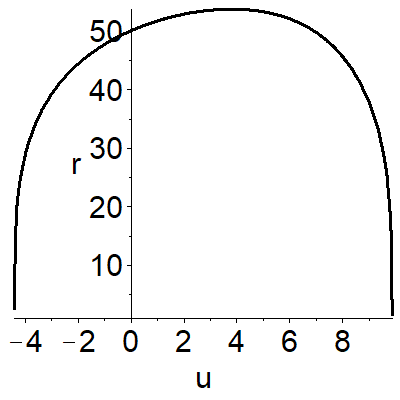} \qquad
\includegraphics[width=0.4\linewidth]{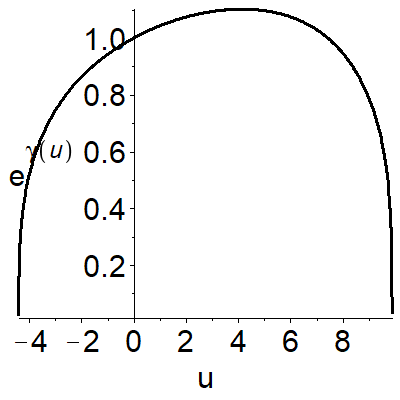} \\
\includegraphics[width=0.4\linewidth]{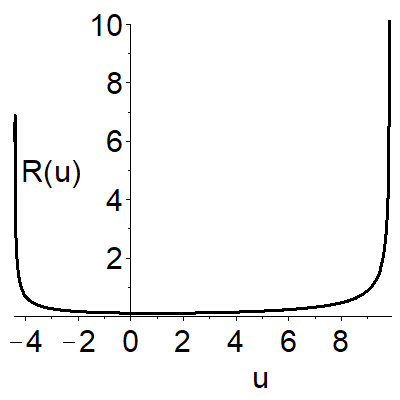}  \qquad
\includegraphics[width=0.4\linewidth]{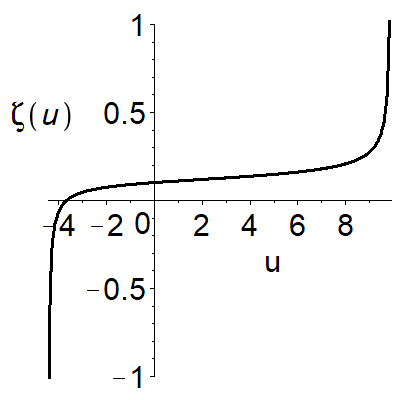} 
\caption{Asymmetric solution for $m_\D=1, \n=3, a=10$, $c=0.2$, $H=0$, $V(\zeta) = m_{\zeta}^2 \zeta^2 /2$, $ m_{\zeta}=0.1$, $r(0)=50$, $r'(0)=2$, $\gamma(0)=0$,  $\gamma'(0)=0.05$, $R(0) \simeq 0.11$ (taken from equation \eqref{cons}), $R'(0) = -0.01$, $\zeta(0) =0.1$, $\zeta'(0) = 0.01$, $u_{\text{min}} \simeq -4.40$, $u_{\text{max}} \simeq 9.89,  r(u_{\text{min}})=r(u_{\text{max}})=0$.}
\label{12}
\end{figure}

Additional conditions could vary continuously leading to a continuum set of inhomogeneous metric functions $\gamma(u), r(u)$ and the scalar field $\zeta(u)$. Some of the distributions are represented in Fig.\ref{12}. We have named these metrics as "flexible" because they smoothly depend on the additional conditions \eqref{cond_2branes}. 

Remark that scalar fields tend to their potential minima in the absence of gravitational interaction. It means that the set of stationary states is quite weak, provided that the potential has finite number of minima (but see \cite{Rubin:2003vj} where a random potential was introduced). It is interaction with the gravity that leads to a continuum set of static solutions and hence, to a continuum set of universes with different properties.

Let us find relation between the $\D$-dimensional Planck mass and 4-dimensional one which is needed to convert units $m_\D=1$ into the physical units. To this end, define
\begin{equation}
R_4 \equiv 12 H^2, \ R_\n \equiv R(u) - \e^{-2 \gamma(u)} R_4,
\end{equation}
see \eqref{Ricci_n_dim_deformed_r}, and require the condition
\begin{equation}
R_\n \gg \e^{-2 \gamma(u)}\, R_4\, 
\end{equation}
which is usually valid for compact extra dimensions treated at low energies, \cite{Bronnikov:2005iz}. Substitution of the Tailor series
\begin{align}
f(R) \simeq & f(R_\n)  + f_{R}(R_\n)\e^{-2 \gamma(u)}R_4 \nonumber \\ & +\frac{1}{2}f_{RR}(R_\n)\e^{-4 \gamma(u)} R_4^2 + ...\, ,
\end{align}
into gravitational part of action \eqref{S0} turns to the effective theory after integration over extra coordinates:
\begin{equation}\label{S_eff}
	S_{eff} = \dfrac{m_{\Pl}^{2}}{2} \int\limits_{M_4} d^4 x \sqrt{|g_4|} \Bigl(a_{eff}R_4^2 + R_4 + c_{eff} \Bigr).
\end{equation}
Here $g_4$ is the determinant of the first quadratic form 
\begin{equation}\label{g4}
ds^2 = dt^2 - \e^{2Ht}\delta_{ij}dx^i dx^j \, ,
\end{equation}
and
\begin{equation} 
\label{m_pl_eff_deformed}
m^2_{\Pl}  = m_\D^{\D-2} \mathcal{V}_{\n-1} \int\limits_{u_{\text{min}}}^{u_{\text{max}}}f_R\bigl(R_{\n}(u)\bigr) \, \e^{2\gamma(u)}\, r^{\n-1} (u)\, du \,,
\end{equation}
where $ \mathcal{V}_{\n-1} =
\int d^{n-1} x \sqrt{|{g}_{n-1}|}=\dfrac{2\pi^{\tfrac{\n}{2}}}{\,\Gamma\left(\tfrac{\n}{2}\right)}$.
Formula \eqref{m_pl_eff_deformed} relates the 4-dimensional Planck mass $m_{\Pl}$ and $\D$-dimensional Planck mass $m_\D$. For the solution shown in the Fig. \ref{11}, $m_{\Pl} \simeq 65.9 m_\D$.

\subsection{Status of static distributions of matter and metric functions}

There are two different aspects concerning the subject of this subsection. The first one relates to the same validity of the classical equations and conditions of the smallness of the quantum effects. Another aspect is the stability of the classical distribution.

\subsubsection*{Quantum perturbations of the extra metric}
Let us estimate the Hubble parameter values for which classical descriptions of our system are valid. It is known that the quantum effects are significant if an action $S< 1$ in the Planck units. The estimation depends on a chosen space region and time interval.  Consider our system acting in a causally connected region of the de-Sitter space of the volume $H^{-3}$  in the time interval $H^{-1}$ when a fluctuation is not stretched enough to leave the chosen region. If the quantum fluctuations are negligibly small, the action satisfies the inequality 
\begin{align}\label{gg1}
S\sim & m_\D^{\n+2}\int\sqrt{|g_\n|} \, d^\n y\, \sqrt{|g_4|}\, d^4 x \, R_4 \nonumber \\ &\sim m_\D^{\n+2} \, l^\n H^{-4} \, 12H^2 \gg 1,
\end{align}
where $l$ is a scale of compact extra space and $H$ is the Hubble parameter.  Hence, classical equations are valid within a causally connected region under condition
\begin{equation}\label{claslim}
    H\ll m_\D^{1+\n/2}\, l^{\n/2},
\end{equation}
which immediately follows from \eqref{gg1}.
On the other side,
$$l\sim H^{-1}$$
according to \eqref{beta_Hubble_ratio}.
Insertion this approximate equality into \eqref{claslim} gives
$$H\ll m_\D. $$
The quantum fluctuations are negligible if this inequality is true.
As is shown in the caption of Fig.\ref{11},
$$m_\D\sim 10^{-2} m_{\Pl}$$
for chosen parameter values.
Hence, classical equations are valid if
\begin{equation}\label{H17}
   H\ll 10^{17} \text{GeV} .
\end{equation}
What about smallness of scalar fields perturbations?

\subsubsection*{Quantum perturbations of the auxiliary field}

The average quantum fluctuations value of the scalar field
\begin{eqnarray}\label{corr}
\sqrt{<\zeta^2>}=\sqrt{\frac{3}{8\pi^2}}\frac{H^2}{ m_{\zeta}} \nonumber
\end{eqnarray}
was obtained earlier, see e.g. \cite{Rey:1986zk, 2021arXiv210908373R}. The quantum fluctuations are negligible if an average value of the classical field $\bar{\zeta}$ satisfies the inequality
$\bar{\zeta}\gg\sqrt{<\zeta^2>}$. This means that the acceptable values of the Hubble parameter $H$ is as follows
\begin{equation}\label{H2}
H\ll m_{\zeta}\bar{\zeta}\sim 0.4\cdot 10^{-5}\cdot 10^{-2}m_{\Pl}\sim 10^{11} \text{GeV}.
\end{equation}
Here numbers are taken from Fig.\ref{11}. The evolution starts at Planckian energies, where quantum fluctuations significantly perturb both the metric and scalar field distribution. The effect of the fluctuations fades away with decreasing energy. Interestingly, the extra-space metric stabilises much earlier, see \eqref{H17}, than the scalar field distribution, see \eqref{H2}.

\subsection{Stability of auxiliary field}

Let us now study the stability at low energies, i.e., $H\simeq 0$ of the solution $\zeta(u)$ more accurately. Recall that this field can be considered as the trial one, provided that its average value is small as compared to the unity, see Fig.\ref{11}. Consider equation
\begin{equation}\label{boxD}
    \square_\D \zeta(t,u) =-m_{\zeta}^2 \ \zeta(t,u)
\end{equation}
to test stability of its solutions. The background metric \eqref{metric_deformed_r} is assumed to be static and $H=0$. By redefining the coordinate $du \rightarrow \e^{\gamma(u)}du$ in \eqref{metric_deformed_r}, equation \eqref{boxD} for solutions of the form $\zeta(t,u)=\zeta_\cc(u)+\delta \zeta (t,u)$ reads 
\begin{align}\label{boxDD}
\e^{-2\gamma(u)} & \left(\delta\ddot{\zeta} + 3H\delta\dot{\zeta} - \delta\zeta'' - \Big(3 \gamma' 
 \right. \nonumber \\ & \left.  
+ \bigl(\n-1\bigr)\dfrac{r'}{r}\Big) \, \delta \zeta' \right)
 =-m_{\zeta}^2\delta\zeta .
\end{align}
It is assumed that the metric and equations of motion have the form represented in the Appendix \eqref{metric_deformed_r_beta} provided that $\beta(u)= \gamma(u)$. 

We will look for solutions in the form $\delta \zeta (t,u)= Z(u)\cdot \e^{i\lambda t}$. In this case, 
the variable changing $Z(u)=\Psi (u) \cdot  \e^{-3\gamma(u)/2} \cdot r^{-(\n-1)/2}(u)$ leads to the Schrodinger-like equation of the form
\begin{equation}\label{Shred}
\Psi''(u) - \Bigl(V(u)-E\Bigr)\Psi(u) = 0,\quad E\equiv\lambda^2
\end{equation}
where
\begin{align}
\label{Potential_Schrod}
&V(u)= \e^{2\gamma(u)} m_{\zeta}^2 \, + \,   \frac32 \gamma '' + \frac94 \gamma'^2 
\nonumber \\ &
+ \frac{\bigl(\n-1\bigr)}{2}\left(\dfrac{3\gamma'r'}{r}+ \dfrac{r''}{r} + \dfrac{\bigl(\n-3\bigr)}{2} \dfrac{{r^\prime}^2}{r^2}\right).
\end{align}
The functions $\gamma(u)$ and $r(u)$ in $V(u)$ from \eqref{Potential_Schrod} are solutions to system \eqref{eq_tt_deformed_r_beta}{--}\eqref{eq_scalar_deformed_r_beta} with constraint equation \eqref{cons_beta} in the case $ \beta(u)= \gamma(u)$. 

\begin{figure}[!th] \centering \includegraphics[width=0.8\linewidth]{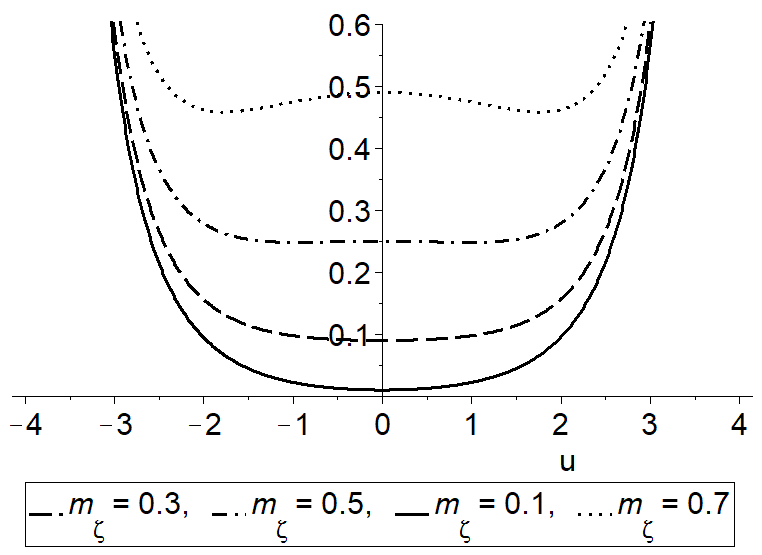} 
\caption{Typical form of potential \eqref{Potential_Schrod} for various masses of the scalar field $m_\zeta$.} \label{13} \end{figure}

As one can see from Fig. \ref{13}, the potential  minimum is positive.
Hence, excitation energy states $E$ are positive as well. It means that the solutions are stable because  $\lambda=\sqrt{E}$ are real.

Stability of the extra space metric was discussed in \cite{Bronnikov:2020tdo}.


\section{Source of countable set of different extra metrics}\label{Source}

In this paper, we assume that our universe was formed at very high energies, somewhere between the inflationary and the Planck scales. Therefore, there must have been a period of descent to the low energies in which we live. The evolution from the Planckian to the inflationary scale is beyond our observation. The natural assumption is that it slowly rolls down in the manner of the inflationary stage \cite{Lindebook}, producing an enormous amount of causally separated regions (pocket universes according to A. Guth, \cite{Guth:1997wk}).

There is another way of producing multiple universes that we will discuss here. Let us start with action \eqref{S0} and a maximally symmetric extra metric with radius depending on time.
\begin{equation}\label{metric_common}
   ds^{2} = dt^2 - \e^{2 \alpha (t)}\delta_{ij}dx^i dx^j - \e^{2 \beta (t)} d\Omega_\n^2 
\end{equation}
Asymptotic solutions \begin{equation} 
\alpha(t)=H_{\text{as}}t, \quad \beta(t)=\beta_{\text{as}}
\end{equation} 
can be found from system \eqref{eq_tt_deformed_r}-\eqref{eq_scalar_deformed_r}.
In this case, the system is strongly simplified:
\begin{align}\label{eq_tt_appr}
 3H^2 f_{R} - \dfrac{f(R)}{2} & = - V_{\text{min}}, \\ 
\label{eq_thetatheta_appr}
- \bigl(\n-1 \bigr)\,m^{2}_{\D}\e^{-2 \beta_{\text{as}}} f_{R} +  \dfrac{f(R)}{2} & = V_{\text{min}}.
\end{align}
Here the Ricci scalar is $R = 12 H^2  + \n \bigl(\n-1 \bigr)\,m^{2}_{\D}\e^{-2 \beta_{\text{as}}}$.
This algebraic system can be resolved in terms of the Hubble parameter $H=H_{\text{as}}$ and the extra-dimensional radius $e^{\beta_{\text{as}}}$. Both values depend on the Lagrangian parameters $a,c$ and a minimum value of the potential $V_{\text{min}}$.
The result can be interpreted as follows. Our manifold starts its evolution at the Planck scale and varies to a stationary asymptotic state described by a constant Hubble parameter and a constant extra-dimensional radius.

Equations \eqref{eq_tt_appr} and \eqref{eq_thetatheta_appr} lead to the relationship
\begin{align}\label{beta_Hubble_ratio}
l\equiv m^{-1}_{\D}\e^{\beta_{\text{as}}} = \sqrt{\dfrac{\n-1}{3}}H_{\text{as}}^{-1},
\end{align}
valid for any form of $f(R)$ and $V(\zeta)$. This means that an extra-dimensional sphere with radius $l$ is uniquely related to the Hubble parameter.
The instabilities of such metrics in the context of the Einstein gravity are discussed in \cite{Contaldi:2004hr}.

The $(4+\n)$-dimensional metric is settled at the de Sitter stage, leading to two known effects - the $3$-dimensional space expands exponentially and it splits into causally separated regions of space - pocket universes. The two processes are permanent. The Hubble parameter $H_{\text{as}}$ should be larger than the inflationary one, $H_I\sim 10^{13}$ GeV, to avoid a possible contradiction with observational data.

Our intermediate goal is achieved: we have the source of the countable set of pocket universes, but another problems remain. Indeed, the properties of all pocket universes are the same, and in addition, it is not clear how they can be shifted down to low energies where the observable physics operates.

To solve these problems, we need to attract quantum fluctuations that are meaningful at high energies. As shown in 
\cite{2021arXiv210908373R}, fluctuations within some of pocket universes are the reason of inhomogeneous extra metric formation. Each fluctuation in any pocket universe tends to a specific static inhomogeneous distribution governed by system \eqref{eq_tt_deformed_r}-\eqref{eq_scalar_deformed_r}, some of them are shown in Figs.\ref{11},\ref{12}.
It breaks the relation \eqref{beta_Hubble_ratio} valid for maximally symmetric metrics
and a part of such regions starts its evolution to low energy states \cite{Petriakova:2022esq}. As a result, we have a large number of low energy states endowed by a variety of inhomogeneous extra metric and non-trivial distributions of the scalar field $\zeta(u)$.

\section{Hierarchy}\label{Hier}

The essence of the Hierarchy problem can be stated as follows. Our experience, based on the inflationary paradigm, indicates that our Universe was formed at high energies ($\ge 10^{14}\,$GeV). It seems reasonable to assume that the values of the physical parameters are on the Planck scale.  On the other hand, the observed values of the physical parameters at low energies vary approximately in the interval $0\div 100\,$GeV. Therefore, there must be an unknown mechanism that reduces the parameter values by more than $12$ orders of magnitude. 

In fact, we already have the necessary tools to create sufficiently small parameters. They are inhomogeneous extra dimensions and some massive matter fields with masses of the order of $m_\D$. Hence, they are invisible to 4D observers but play a key role when their interaction with other fields is taken into account.

Note that the scalar fields appear naturally already in the standard 4-dimensional gravity \cite{1992PhR...215..203M,Alvarenga:2023oep} as scalar functions describing scalar perturbations of the metric around the background metric. Evidently, the $\D$-dimensional gravity provides more opportunities. 
For example, small perturbations $\delta\gamma(X)\ll 1$, $\delta r(X)\ll r(X)$ of the scalar  functions $\gamma(X)$ and $r(X)$ in \eqref{metric_deformed_r} can be considered as trial scalar fields acting in the background spacetime. 
This means that these scalars are perspective candidates to the role of small parameters necessary to solve the Hierarchy problem.  Here we use the field $\zeta$ for this purpose.

\subsection{Source of small parameters}\label{ssp}

The smallness of a Lagrangian parameters can be attributed to an inhomogeneous extra metric, as shown in \cite{Rubin:2020pqu}. This procedure is effective when the number of such parameters is not very large. A large number of observable parameters, as in the Standard Model, complicates the procedure. This is a reason to involve another degrees of freedom like scalar fields at high energies, the scalar perturbations of extra metric, radion, dilaton, the Ricci scalars for example. Here we consider a scalar field $\zeta$ with action \eqref{S0} and treat it as the background field, see Introduction

Now let us extend general relation \eqref{l4}, accounting dependence of effective 4-dimensional physical parameters $\lambda_4$ on the field $\zeta$
\begin{equation}\label{l4zeta}
    \lambda_{4}=\mathcal{Z} (g_\n(y),\zeta(y),\lambda_{4+\n})
\end{equation}
after integration out extra coordinates in a $\D$-dim action. Here $\lambda_4 \ll 1$ that is true for known physical parameters while $\lambda_{4+\n}\sim 1$ according to our agreement. Function $\zeta(u)$ is classical solution to the equations of motion \eqref{boxD}. 

As discussed above and in \cite{Petriakova:2022esq}, the pure de Sitter space formed at the scale larger than the scale $\sim 10^{14}$ GeV produces countable set of pocket universes. 
Each of them contains the background  field $\zeta(u)$, which is the result of asymptotic evolution of the field $\zeta(t,u)$ caused by an initial quantum fluctuation in a particular pocket universe. Its average values $\bar{\zeta}$ acquire random values in the interval $0\leq|\bar{\zeta}|\lesssim 1$ in a variety of the pocket universes. In particular, some of pocket universes contain small values  $\bar{\zeta}_{small}\lll 1$ as in Fig.\ref{11}, right panel. 
It will be shown below that they could stand by a small parameter and significantly facilitate solution to the Hierarchy enigma.

To proceed, we must come into agreement on what is meant by the "natural values" of the Lagrangian parameters $g_l$.
It is definitely matter of taste, but a restriction $10^{-2}<g_l< 10^{2}$ looks as appropriate choice. So, we will suppose that ''natural'' range for a parameter value deviates from unity not more than two orders of magnitude. More definitely, any physical parameter $g_l$ of a dimensionality $l\neq 0$ can be represented in the form $g_l=(C_l\, m_\D)^l$ where $m_\D$ is the energy scale, $\D$-dimensional Planck mass in our case. It is the dimensionless parameter $C_l$ which is assumed to be "natural" if $C_l\sim 10^{\pm 2}$. Therefore, natural values of physical parameter $g_l$ vary within the interval $(10^{-2l}\div 10^{+2l})$. For example, let the dimensionality of a parameter $\lambda$ is $[m_\D]^{2}$. It means that we are free to choose initial value of $\lambda$ in a wide interval $(10^{-4}\div 10^{4})m_\D^2$ and still consider the choice as natural.


\subsection{The shift of the Higgs field down to low energies}

In this section we discuss the Hierarchy problem using the Higgs field as an example. Within the framework 
of our approach outlined in the Introduction, we assume that the physics of the Higgs field together with the background field are formed at the Planck scale.

Suppose that a form of the Higgs action at the Planck scale is the same as at the electroweak scale but the Lagrangian parameters depend on the background field $\zeta$ (see discussion between formulas \eqref{S0} and \eqref{eqMgravity} and the Introduction), 
\begin{eqnarray}\label{SH}
S_{\HH_\text{P}} &=&
\frac12 \int d^{\D} X \sqrt{|g_{\D}|} \, \Bigl(\partial^{\M} {H_\text{P}}^\dagger \partial_{\M} H_\text{P} 
\nonumber \\ &&
+ \nu(\zeta) {H_\text{P}}^\dagger H_\text{P} - \lambda(\zeta)\bigl({H_\text{P}}^\dagger H_\text{P}\bigr)^2 \Bigr), 
\end{eqnarray}
where $ \nu\bigl(\zeta(u)\bigr), \lambda\bigl(\zeta(u)\bigr)  > 0$ are arbitrary functions, $\zeta(u)$ is a solution to equation \eqref{eqwr} and $H_\text{P}$ is a proto Higgs field. 

All numerical values in the Lagrangian \eqref{SH} are of the order of the unity, in $m_D$ units. Our aim is to show that the parameters can be reduced in many orders of magnitude by appropriate choice of the inhomogeneous metric and the background field $\zeta$.

Variation of \eqref{SH} with respect to $H_\text{P}$ gives
\begin{equation}\label{boxHP}
\square_\D H_\text{P}=\nu(\zeta) H_\text{P} - 2 \lambda(\zeta) \bigl({H_\text{P}}^\dagger H_\text{P}\bigr)H_\text{P}.
\end{equation}

Equations of motion for the background field is as follows
\begin{equation}\label{eqwr}
\square_\n \zeta(u) \simeq -{m_{\zeta}}^2 \zeta(u). 
\end{equation}
We assume that the  background field dominates over terms containing the field $H_\text{P}$.
The proto Higgs field is seeking as
\begin{equation}\label{Hxu}
H_\text{P} = h(x) \ {U}(u), 
\end{equation}
where $h(x)$ is 2-components column acting in the fundamental representation of $SU(2)$.
Our immediate aim is to find the distribution of the field $H_\text{P}$ over the extra coordinates ruled by the scalar function $U(u)$ by solving system \eqref{boxHP} and \eqref{eqwr}.

\begin{figure}[!th]
\centering
\includegraphics[width=0.5\linewidth]{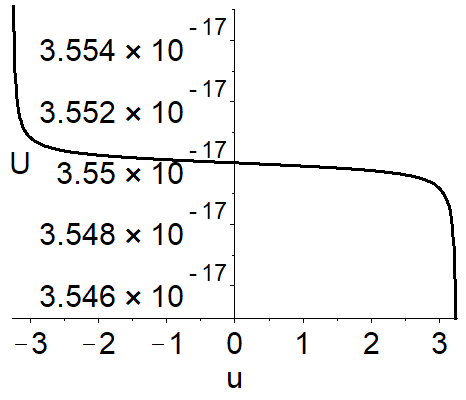}
\caption{The solution of the equation \eqref{boxU} for $\nu(\zeta)=\mu^2 \zeta^2$, $\mu=2$, $\lambda = 200$, $v=1$, $U(0) =3.55 \cdot 10^{-17}$, $U'(0) =-1 \cdot 10^{-21}$  on the background shown in Figure \ref{11}. }
\label{15}
\end{figure}

The inhomogeneities of the field $h(x)$ are important at the low energies, but they are exponentially stretched during the first, de Sitter-like stage, see section \ref{Source}, so that $h(x)=v= \text{const}$ with great accuracy.
It means that 
\begin{equation}\label{Hv}
h(x) = \frac{1}{\sqrt{2}} 
\begin{pmatrix}
0 \\ v+\rho(x)
\end{pmatrix}
\simeq 
\frac{1}{\sqrt{2}} 
\begin{pmatrix}
0 \\ v
\end{pmatrix}
.
\end{equation}
Therefore, approximation \eqref{Hv} transforms
Eq. \eqref{boxHP} in the following way
\begin{equation}\label{boxU}
\square_\n  U(u) =\nu(\zeta)  U(u) - \lambda(\zeta) \, v^2 U^3(u), 
\end{equation}
with yet unknown parameter $v.$
There is infinite set of solutions to this equation because the internal space metric and the background field $\zeta$ are inhomogeneously distributed over the coordinate $u$. We have chosen one among a continuum set of solutions valid at small energy scale (see Figure \ref{15}). 

Knowledge of solutions to equations \eqref{boxU} and \eqref{eqwr} permits us to integrate out the internal coordinates and reduce action \eqref{SH} to the 4-dimensional form
\begin{align}\label{SHH}
S_{\HH} =& \frac{\mathcal{V}_{\n-1}}{2}\int d^{4} x  \sqrt{|\tilde{g}_{4}|} \int\limits_{u_{\text{min}}}^{u_{\text{max}}}  \biggl(\e^{-2\gamma(u)}  U^2(u) \tilde{g}^{i j} \partial_{i} h^\dagger\partial_{j} h \nonumber \\ 
& + \Bigl(-(\partial_u U)^2+ \nu(\zeta) \, U^2(u)\Bigr) h^\dagger h \nonumber \\ 
&  -  \lambda(\zeta)\,  U^4(u) \bigl(h^\dagger h\bigr)^2 \biggr) \e^{4\gamma(u)}r^{\n-1}(u) \, du
\end{align}
after substitution \eqref{Hxu} into \eqref{SH}. 
To study this action at low energies, we choose the Minkowski metric
\begin{equation}
d\tilde s^2 = \eta_{i j} d x^i d x^j.
\end{equation}
with the Hubble parameter   $H=0$.
Definition of parameters 
\begin{align}
\label{Kh}
K_h &= \mathcal{V}_{\n-1} \int\limits_{u_{\text{min}}}^{u_{\text{max}}}  U^2(u) \, \e^{2\gamma(u)}r^{\n-1}(u)\, du,  
\\
\label{m_rho}
m^2_h & =\mathcal{V}_{\n-1} \int \limits_{u_{\text{min}}}^{u_{\text{max}}} \Bigl(-(\partial_u U)^2  \nonumber \\ &
+\nu(\zeta) \, U^2(u)\Bigr)\e^{4\gamma(u)}r^{\n-1}(u) \, du,  \\ \label{lh}
\lambda_h &= \mathcal{V}_{\n-1} \int\limits_{u_{\text{min}}}^{u_{\text{max}}} \lambda (\zeta)\,U^4(u)\, \e^{4\gamma(u)}r^{\n-1}(u) \, du,
\end{align}
depending on the internal coordinate $u.$ The variable changing \begin{equation}\label{Hh}
H_0={h\sqrt{K_h}}
\end{equation}
leads to the 4-dimensional effective Higgs Lagrangian
\begin{align}\label{SHHH}
S_{\HH} =& \frac{1}{2}\int d^{4} x  \sqrt{|\tilde{g}_{4}|} \int\limits_{u_{\text{min}}}^{u_{\text{max}}}  \biggl(\partial_{i} H_0^\dagger\partial^{i} H_0 
\nonumber \\ & 
+  m_H^2 H_0^\dagger H_0 -  \lambda_H\bigl(H_0^\dagger H_0\bigr)^2 \biggr)\, du, \\
& m_H^2\equiv\frac{m_h^2}{K_h},\quad \lambda_H\equiv\frac{\lambda_h}{K_h^2}.\label{mlH}
\end{align}
Here $H_0$ is the observable Higgs field. 
Experimentally measured parameters are
the Higgs boson mass  and its vacuum average
\begin{equation}\label{obs}
m_{Higgs}=125 \, \text{GeV},\quad v_{Higgs}=246\, \text{GeV}\,.
\end{equation}
according to \cite{Workman:2022ynf}.
They are related to the parameters $m_H$ and $\lambda_H$ of the effective Higgs action \eqref{SHHH} as follows 
\begin{equation}\label{mH}
m_{H} =m_{Higgs} /\sqrt{2}=88.6\,GeV \simeq 10^{-17} m_{\Pl}, \end{equation}
and \begin{equation}\label{lH}
\lambda_{H}=(m_{Higgs}/v_{Higgs})^2/2  \simeq 0.13, \end{equation} 

Remark that all formulas contain the function $U(u)$, the solution to equation \eqref{boxU} whith yet unknown constant $v$. Interestingly, the Lagrangian structure \eqref{SH} allows to avoid the determination of this constant. Indeed, a solution to equation \eqref{boxU} can be found for the function 
$$\tilde{U}(u) = vU(u)$$  
because Eq. \eqref{boxU} for the function $\tilde{U}(u)$ does not contain unknown parameter $v$ in this case. Moreover, the substitution of $U=\tilde{U}(u)/v$ into expressions \eqref{Kh}, \eqref{m_rho} and \eqref{lh} 
gives
\begin{align}\label{UtU}
&K_h[U]= \frac{K_h[\tilde{U}]}{v^2} , \quad m_h^2[U]=  \frac{m_h^2[\tilde{U}]}{v^2} , \nonumber \\ & \lambda_h[U]=  \frac{\lambda_h[\tilde{U}]}{v^4}.
\end{align}
and hence, the parameters $m_H,\lambda_H$ in \eqref{mlH}  do not depend $v$. This value appears also in relation 
\begin{equation}\label{vcalc}
v=v_{Higgs}/\sqrt{K_h[U]}
\end{equation}
followed from \eqref{Hh} and the substitution $H\to v,\, h\to v_{Higgs}$. 
Luckily, this relation does not depend on $v$ as well. After taking into account first equality in \eqref{UtU}, we obtain additional restriction to the function $\tilde{U}$
\begin{equation}\label{vcalc1}
1=v_{Higgs}/\sqrt{K_h[\tilde{U}]}.
\end{equation}

All formulas written above are valid for an arbitrary functions $\nu(\zeta), \lambda(\zeta)$. To proceed, we choose their specific form as $$\nu(\zeta) = \mu^2 \zeta^2, \quad \lambda=\lambda_0=const.$$
Calculations show that the relations (\ref{mH}), (\ref{lH}) and (\ref{vcalc1}) are met under the following conditions: $ \mu =2, \ \lambda = 200.$ 
There is one Lagrangian parameter the value of which could be suspected as unnaturally large - $\lambda$. More detailed analysis shows that this value is still ''natural'' in the agreement discussed in the beginning of Sect.\ref{Hier}.
Indeed, the dimensionality of $\lambda$ is $[m_\D]^{2}$ for $\n=3$. Hence, natural interval for $\lambda$ is $10^{\pm 4}$ and our choice $\lambda =\text{const} = 200$ looks acceptable. 
The origin of small parameters has been discussed earlier, see the beginning of Section \ref{Hier}. The quantum fluctuations produce a variety of the field amplitudes in the countable set of the pocket universes. A part of them contains small amplitudes of $\zeta$ and $U$.
The fields values of the order of $\zeta \sim 10^{-4}$ (see Fig.\ref{11}) and $U\sim 10^{-16}$ (see Fig.\ref{15})  suit for relations (\ref{mH}), (\ref{lH}) and (\ref{vcalc1}). 

In this paper, we do not discuss the cosmological constant value because too many effects have to be taken into account. As the example, estimate the energy density stored in the auxiliary field $\zeta$ 
\begin{align}\label{infdens}
\varepsilon & \sim \frac{m_{\D}^{\D-2}}{2} \mathcal{V}_{\n-1} \int \limits_{u_{\text{min}}}^{u_{\text{max}}}  {m_{\zeta}}^2\zeta^2(u) \e^{4\gamma(u)}r^{\n-1}(u) \, du \nonumber \\ & \simeq 6.4 \cdot 10^{-9}m_\D^4\sim 3.4 \cdot 10^{-16} m_{\Pl}^4 \, . 
\end{align}
Other effects should compensate this value that is the essence of the fine tuning problem \cite{Weinberg:1988cp}.

\section{Conclusion}
Solving the Hierarchy puzzle can be successful if we have a set of small parameters. In our paper we suggest how such small parameters arise and apply them to the Higgs field, the important part of the Standard Model. 

It is assumed that our Universe was nucleated at high energies. The process of reduction from high to low energies is as follows. A $\D$-dimensional manifold evolves from the Planck scale to a stage characterised by the de-Sitter metric of 4D space and a static metric of $\n$-dimensional extra space. The 4-dimensional space exponentially expands in its 3 dimensions and disconnected space regions - pocket universes - are continuously created. Quantum fluctuations  independently disturb the extra dimensional metric within these universes, causing inflationary processes within them.  Some of these universes tend asymptotically to the low energy states in which we operate. The effective parameter reduction applied to the Higgs sector of the Standard Model is explained by the presence of small-amplitude static distributions of a background scalar field $\zeta$ in a fraction of these universes.

Background scalar field is necessary element for the successful realization of the idea. There is continuum set of such field distributions acting in the inhomogeneous metric of extra dimensions within the pocket universes. The distributions caused by the quantum fluctuations vary from zero  to the Planck values. The amplitudes of some of them are extremely small which is the reason for the smallness of the observable physical parameters. 

The Higgs parameters are fitted with good accuracy, which is not mandatory because other factors affect the result. The aim here is to show that an effective reduction of the physical parameters from high to low energy can be achieved.

A flexible inhomogeneous extra space is the basis of our approach to solve the Hierarchy problem. The flexible space metric was also applied to the problem of the observed baryon asymmetry \cite{Nikulin:2020nub,Kirillov:2012gy}, which deserves a several words. On the one hand, the appearance of the asymmetry implies that the baryon charge was not conserved from the beginning. On the other hand, the baryon charge is conserved with great accuracy at the present time. The flexible extra dimensions allow us to resolve this contradiction. In our previous works we have shown the ability of the flexible metrics, but the introduction of background fields opens up new possibilities.


\subsection*{Appendix}
Here we provide the necessary expressions for working with an inhomogeneous n-dimensional extra metric of a more general kind
\begin{align}\label{metric_deformed_r_beta}
ds^{2} & = \e^{2\gamma(u)}\Bigl(dt^2 - \e^{2Ht}\delta_{ij}dx^i dx^j\Bigr) \nonumber \\ & - \e^{2\beta(u)}du^2 - r^2(u)\,d\Omega_{\n-1}^2 \, , \quad i,j=\overline{1,3}
\end{align}
The Ricci scalar
\begin{align}\label{Ricci_n_dim_deformed_r_beta}
R(u) &= 12 H^2 \e^{-2 \gamma(u)} - \Biggl( 8 \gamma'' + 20 {\gamma'}^2  - 8 \gamma' \beta' 
\nonumber \\ &
  + \bigl(\n-1 \bigr)\biggl( \dfrac{2 r^{\prime\prime}}{r} + \dfrac{8 \gamma' r^{\prime}}{r} -  \dfrac{2 \beta' r^{\prime}}{r} 
  \nonumber \\ & + \dfrac{\bigl(\n-2 \bigr)}{r^2} \Bigl(r'^2 - \e^{2\beta(u)}\Bigr) \biggr) \Biggr)\e^{-2\beta(u)} 
\end{align}
does not depend on time. The notation used are ${}' \equiv d / du$ and ${}'' \equiv d^2 / du^2$ respectively.
Equations \eqref{eqMgravity} for $(tt)=...=(x^3 x^3)$, $(uu)$ and $(x^5 x^5)=...=(x^{\D-1} x^{\D-1})${--}components and \eqref{eqMscalarfield} become as follows
\begin{align}\label{eq_tt_deformed_r_beta}
& \Biggl( {R'}^2 f_{RRR} +\Big(R'' +3 \gamma' R' - \beta' R'  \nonumber \\  &  + \bigl(\n-1 \bigr) \frac{r'}{r} R' \Big)f_{RR} - \Big( \gamma'' +4{\gamma'}^2 - \beta' \gamma' 
\nonumber \\  & 
+ \bigl(\n-1 \bigr)\frac{\gamma' r'}{r} \Big)  f_{R} \Biggr) \e^{-2\beta(u)}  +  \frac{3H^2}{\e^{2 \gamma(u)}} f_{R} 
\nonumber \\  & 
- \dfrac{f(R)}{2} = - \dfrac{\zeta'^2}{2} \, \e^{-2\beta(u)}   - V\bigl( \zeta \bigr), 
\end{align}
\begin{align}
\label{eq_thetatheta_deformed_r_beta}
& \Biggl(\Big( 4\gamma'R' + \bigl(\n-1 \bigr) \dfrac{r'}{r} R' \Big)\,f_{RR} - \Big( 4 \gamma'' + 4{\gamma'}^2 
\nonumber \\  & 
- 4 \gamma' \beta' + \bigl(\n-1 \bigr) \dfrac{r''}{r} - \bigl(\n-1 \bigr) \frac{\beta' r'}{r}\Big) \, f_R \Biggr)\e^{-2\beta(u)} \nonumber  \\ &  - \dfrac{f(R)}{2} =    \dfrac{{\zeta'}^2}{2} \, \e^{-2\beta(u)} - V\bigl(\zeta \bigr) \,, 
\end{align}
\begin{align}
\label{eq_phiphi_deformed_r_beta}
& \Biggl( {R'}^2 f_{RRR} \, + \biggr( R'' + 4\gamma' R' + \bigl(\n - 2\bigr) \dfrac{r'}{r}\,R' \nonumber
\\  & 
- \beta' R'  \biggl) f_{RR} - \biggl(\dfrac{r^{\prime\prime}}{r} + \frac{4\gamma' r'}{r} - \frac{\beta' r'}{r} \nonumber \\  & + \dfrac{\bigl(\n-2 \bigr)}{r^2} \left(r'^2 - \e^{2\beta(u)}\right) \biggr) f_R  \Biggr)\e^{-2\beta(u)} \nonumber \\  & - \, \dfrac{f(R)}{2} = -  \dfrac{{\zeta'}^2}{2}\, \e^{-2\beta(u)} - V\bigl(\zeta \bigr) \, , 
\end{align}
\begin{align}
\label{eq_scalar_deformed_r_beta}
& \left( \zeta'' + \left(4 \gamma' - \beta' + \bigl(\n-1\bigr)\dfrac{r'}{r}\right) \, \zeta' \right)\e^{-2\beta(u)} \nonumber \\ & - V^{\prime}_{\zeta} =0.
\end{align} 
with the constraint equation \begin{align}\label{cons_beta}
& \left(\biggr( 8\gamma'R' + 2\bigl(\n-1\bigr)\dfrac{r'}{r}R' \biggl) f_{RR} + \Biggl(12 {\gamma'}^2 
\right. \nonumber \\ & 
 +\bigl(\n-1\bigr) \biggl(\dfrac{8\gamma' r'}{r} + \dfrac{\bigl(\n-2 \bigr)}{r^2} \left(r'^2 - \e^{2\beta(u)}\right) \biggr) 
\nonumber \\ & \left.
+R \Biggr)f_R\right)\e^{-2\beta(u)} - 12H^2\e^{-2 \gamma(u)} f_R\, 
\nonumber \\ &
- f(R) =  {\zeta'}^2 \e^{-2\beta(u)} - 2 V\bigl( \zeta \bigr)	
\end{align}
containing first-order derivatives from the combination $2\cdot$\eqref{eq_thetatheta_deformed_r_beta}$-f_R \cdot$\eqref{Ricci_n_dim_deformed_r_beta}. 

\section{Acknowledgments}

The work of SGR was funded by the Ministry of Science and Higher Education of the Russian Federation, Project "New Phenomena in Particle Physics and the Early Universe" FSWU-2023-0073
and the Kazan Federal University Strategic Academic Leadership Program. The work of AAP was funded by the development program of the Regional Scientific and Educational Mathematical Center of the Volga Federal District, agreement N 75-02-2022-882. P.P. is grateful to the Foundation for the Advancement of Theoretical Physics and Mathematics “BASIS” for financial support, Grant No. 22-1-5-114-1.

\printbibliography[title={References}, heading=bibintoc]

\end{document}